\documentclass[12pt]{iopart}

\usepackage{graphicx}
\usepackage{here}

\usepackage{stackrel}
\usepackage{amssymb,accents,latexsym,mathrsfs,bm}

\newcommand{\SU}{{\textrm {SU}}}
\newcommand{\U}{{\textrm {U}}}

\newcommand{\cA}{{\cal A}}

\newcommand{\cW}{{\cal W}} 
 
\newcommand{\mcA}{{\mathcal A}}
\newcommand{\mcV}{{\mathcal V}}
\newcommand{\mbbV}{{\mathbb{V}}}
\newcommand{\mbhV}{{\mathscr{V}}}
\newcommand{\mcI}{{\mathcal I}}
\newcommand{\mcT}{{\mathcal T}}

\newcommand{\PPFF}{{\scriptsize\textrm{PF}}}
\newcommand{\cons}{{\textrm{\scriptsize cons}}}
\newcommand{\cov}{{\textrm{\scriptsize cov}}}
\newcommand{\BZ}{{\mathrm{BZ}}}
\newcommand{\MeV}{{\textrm{MeV}}}

\newcommand{\eem}{{\textrm{\scriptsize em}}}
\newcommand{\bbar}{{\textrm{\scriptsize bar}}}
\newcommand{\iiso}{{\textrm{\scriptsize iso}}}
\newcommand{\V}{{\textrm{\scriptsize V}}}

\newcommand{\A}{{\textrm{\scriptsize A}}}
\newcommand{\LL}{{\textrm{\scriptsize L}}}
\newcommand{\RR}{{\textrm{\scriptsize R}}}
\newcommand{\gauge}{{\textrm{\scriptsize gauge}}}

\newcommand{\WZW}{{\textrm{\scriptsize WZW}}}
\newcommand{\YM}{{\textrm{\scriptsize YM}}}

\begin{document}

\title[Constitutive relations of a chiral hadronic fluid]{Constitutive relations of a chiral hadronic fluid}

\author{Juan L. Ma\~nes$^{1}$, Eugenio~Meg\'{\i}as$^{2}$, Manuel Valle$^{1}$ and \\
Miguel \'A. V\'azquez-Mozo$^{4}$}
\address{
$^{1}$ Departamento de F\'{\i}sica, Universidad del Pa\'{\i}s Vasco UPV/EHU, Apartado 644, 48080 Bilbao, Spain  \\

$^{2}$ Departamento de F\'{\i}sica At\'omica, Molecular y Nuclear and Instituto Carlos I de F\'{\i}sica Te\'orica y Computacional, Universidad de Granada, \\ 
Avenida de Fuente Nueva s/n, 18071 Granada, Spain \\

$^{3}$ Departamento de F\'{\i}sica Fundamental, Universidad de Salamanca, Plaza de la Merced s/n, 37008 Salamanca, Spain
}
\ead{wmpmapaj@lg.ehu.es, emegias@ugr.es, manuel.valle@ehu.es, Miguel.Vazquez-Mozo@cern.ch}

\vspace{10pt}

\begin{abstract}
We study the role of non-abelian anomalies in relativistic fluids. To this end, we compute the local functional that solves the anomaly equations, and obtain analytical expressions for the covariant currents and the Bardeen-Zumino terms. We particularize these results to a background with two flavors, and consider the cases of unbroken and broken chiral symmetry. Finally, we provide explicit results for the constitutive relations of chiral nuclear matter interacting with external electromagnetic fields and in presence of chiral imbalance. We emphasize the non-dissipative nature of the chiral electric effect.
\end{abstract}

%

\section{Introduction}
\label{sec:introduction}

Hydrodynamics is an effective description of out-of-equilibrium systems in which the mean free path of particles is much shorter than any macroscopic time or length scale of the system~\cite{Kovtun:2012rj}. The equations of motion are the conservation laws of the energy-momentum tensor and charged currents, and they are supplemented by the so-called constitutive relations, i.e. expressions of these quantities in terms of fluid variables and organized in a derivative expansion. Quantum anomalies are associated with very robust mathematical properties of gauge fields at the non perturbative level~\cite{Zumino:1983ew,AlvarezGaume:1983}. In the presence of anomalies the currents are no longer conserved, and this has important effects in the hydrodynamic description of relativistic fluids. In addition to the ideal hydrodynamical contributions, there are extra terms in the constitutive relations which lead to dissipative and anomalous effects, i.e. for the charged currents~$\langle J^\mu \rangle =  n u^\mu +  \langle \delta J^\mu \rangle_{\textrm{\scriptsize diss \& anom}}$. In the presence of external electromagnetic fields or vortices in the fluid, parity is broken and some tensor structures appear in the constitutive relations associated to time reversal transport. This is the case of the {\it chiral magnetic effect} (CME), which is responsible for the generation of an electric current induced by an external magnetic field~\cite{Fukushima:2008xe}, and the {\it chiral vortical effect} (CVE) in which the electric current is induced by a vortex~\cite{Son:2009tf}, i.e.  $\langle \delta J^\mu\rangle_{\textrm{\scriptsize anom}} = \sigma_B {\mathcal B}^\mu + \sigma_V \omega^\mu$. The corresponding susceptibilities, $\sigma_B$ and $\sigma_V$, are related to non-dissipative phenomena as they do not contribute to entropy production. These coefficients have been computed in a wide variety of methods, including kinetic theory~\cite{Stephanov:2012ki,Huang:2017tsq}, Kubo formulae~\cite{Kharzeev:2009pj,Landsteiner:2011cp,Landsteiner:2012kd} and fluid/gravity correspondence~\cite{Bhattacharyya:2008jc,Erdmenger:2008rm,Landsteiner:2011iq,Megias:2013joa}. 

It has been recently proposed a new formalism to obtain the non-dissipative part of the anomalous constitutive relations based on the existence of an equilibrium partition function in a generic stationary background~\cite{Banerjee:2012iz,Jensen:2013vta}. There are several strategies to compute the effective action, including the solution of the anomaly equations~\cite{Manes:2018mth}, as well as differential geometry methods~\cite{Jensen:2013kka,Manes:2018llx,Manes:2019cqm}. Some applications of these techniques to the physics of anomalous fluids in thermal equilibrium have been presented in e.g. Refs.~\cite{Jensen:2012kj,Haehl:2013hoa}. In this work we will use this formalism to study the anomalous contributions to the constitutive relations in non-abelian theories. We also extend this analysis to study the hydrodynamics in presence of spontaneous symmetry breaking. In this case the equilibrium partition function is computed from the Wess-Zumino-Witten (WZW) functional, thus describing the low-energy interaction of Nambu-Goldstone (NG) bosons with external gauge fields. We apply these results to the analysis of chiral nuclear matter fluids in the presence of baryon, isospin and axial chemical potential. Our results show the existence of the chiral electric effect (CEE) first predicted in~\cite{Neiman:2011mj}, and confirm its non-dissipative nature.

\section{Equilibrium partition function formalism and hydrodynamics}
\label{sec:equilibrium_partition_function}

We will present in this section the main ingredients of the equilibrium partition function formalism relevant to compute the anomalous contributions to the constitutive relations~\cite{Banerjee:2012iz,Jensen:2013kka,Bhattacharyya:2013lha,Megias:2014mba}.

\subsection{Equilibrium partition function}
\label{subsec:equilibrium_partition_function}

Let us consider a relativistic invariant quantum field theory with a time independent $\U(1)$ gauge connection on the manifold~\footnote{For simplicity, we will be restricted to abelian theories in this section, but we will generalize this analysis to the non-abelian case in Sec.~\ref{sec:non_abelian}.}
\begin{eqnarray}
ds^2 &=& G_{\mu\nu} dx^\mu dx^\nu  = - e^{2\sigma(\vec{x})}(dt + a_i(\vec{x}) dx^i)^2 + g_{ij}(\vec{x}) dx^i dx^j \,,  \label{eq:metric} \\
{\cal A} &=& {\cal A}_0(\vec{x}) dx^0 + {\cal A}_i(\vec{x}) dx^i \,.
\end{eqnarray}
The partition function of the system is 
\begin{equation}
Z = \Tr \, e^{-\frac{H-\mu_0 Q}{T_0}} \,,
\end{equation}
where $H$ is the Hamiltonian, $Q$ is the charge associated to the gauge connection, while $T_0$ and $\mu_0$ are the temperature and chemical potential at equilibrium. The dependence of $Z$ on the fields $\{\sigma, g_{ij}, a_i, \cal A_\mu\}$ should be consistent with: i) three-dimensional diffeomorphism invariance; ii) Kaluza-Klein (KK) invariance, i.e. $t \to t + \phi(\vec{x}) \,, \; \vec{x} \to \vec{x}$; and iii) $\U(1)$ time-independent gauge invariance (up to an anomaly). From the partition function of the system, one can compute the energy-momentum tensor and $\U(1)$ charged current by performing the appropriate $t$-independent variations, i.e.
\begin{equation}
\delta \log Z = \frac{1}{T_0}\int d^3x \sqrt{g_3} \, e^{\sigma} \left( -\frac{1}{2} T_{\mu\nu} \delta g^{\mu\nu} + J^\mu \delta {\cal A}_\mu \right) \,,
\end{equation}
where $g_3 = \det(g_{ij})$. The KK invariance of the partition function demands that $\log Z$ depends on the gauge fields through the following invariant combinations
\begin{equation}
A_0 = \cA_0 \,, \qquad A_i = \cA_i - a_i \cA_0  \,. \label{eq:A_KK}
\end{equation}
For a general dependence $\log Z = {\mathcal W}(e^\sigma, A_0, a_i, A_i, g^{ij},T_0,\mu_0)$, one gets the consistent currents and energy-momentum tensor~\cite{Banerjee:2012iz}
\begin{eqnarray}
&&\hspace{-2cm}  \langle J_0 \rangle_\cons = -\frac{T_0 e^{\sigma}}{\sqrt{g_3}}\frac{\delta \mathcal{W}}{\delta A_0} \,,  \quad \langle J^i \rangle_\cons = \frac{T_0 e^{-\sigma}}{\sqrt{g_3}}\frac{\delta \mathcal{W}}{\delta A_i}  \,, \quad \langle T^{ij} \rangle =  -\frac{2 T_0 e^{-\sigma}}{\sqrt{g_3}} g^{ik} g^{jl}  \frac{\delta \mathcal{W}}{\delta g^{kl}} \,,  \label{eq:Jcr} \\
&&\hspace{-2cm} \langle T_{00} \rangle =-\frac{T_0 e^{\sigma}}{\sqrt{g_3}}\frac{\delta \cW}{\delta \sigma} \,, \quad\hspace{0.4cm} \langle T_0^{\;i} \rangle = \frac{T_0 e^{-\sigma}}{\sqrt{g_3}}\left(\frac{\delta \mathcal{W}}{\delta a_i} - A_0 \frac{\delta \cW}{\delta A_i}\right) \,, \label{eq:Tcr}
\end{eqnarray}
so that ${\mathcal W}$ is a generating functional for the hydrodynamic constitutive relations.

\subsection{Derivative expansion}
\label{subsec:derivative_expansion}

Let us study the properties of the partition function in a derivative expansion. The most general equilibrium partition function up to zeroth order in derivatives is~\cite{Banerjee:2012iz}
\begin{equation}
\log Z = {\cal W}_{(0)} = \frac{1}{T_0} \int d^3x \sqrt{g_3} \, e^\sigma P(e^{-\sigma} T_0, e^{-\sigma} A_0)  \,,
\end{equation}
where $P$ is an arbitrary function of two variables. Then the constitutive relations can be written as 
\begin{eqnarray}
&& \langle J^0 \rangle = e^{-\sigma} \partial_b P \,, \qquad \langle J^i \rangle = 0 \,, \label{eq:crJ} \\
&&\hspace{-0.1cm} \langle T^{ij} \rangle = P g^{ij} \,,  \qquad\hspace{0.3cm} \langle T_{00}\rangle = e^{2\sigma}(P - a \partial_a P - b \partial_b P) \,, \qquad \langle T_0^{\;i} \rangle = 0 \,,  \label{eq:crT}
\end{eqnarray}
where we have used the notation $a \equiv e^{-\sigma} T_0$ and $b \equiv e^{-\sigma} A_0$. By comparison with the hydrodynamic constitutive relations of a perfect fluid (PF)
\begin{equation}
\langle J^\mu \rangle_{\PPFF} = n u^\mu  \,, \qquad \langle T^{\mu\nu} \rangle_{\PPFF}  = (\varepsilon + {\cal P}) u^\mu u^\nu + {\cal P} g^{\mu\nu} \,,
\end{equation}
where $\varepsilon$ is the energy density, ${\cal P}$ the pressure, $n$ the charge density and $u^\mu$ the local fluid velocity, one gets
\begin{equation}
\hspace{-1cm} u^\mu = e^{-\sigma}(1, 0, \dots, 0) \,, \quad {\cal P} = P \,, \quad \varepsilon = -P + a \partial_a P + b \partial_b P \,, \quad n = \partial_b P \,. \label{eq:umu_P}
\end{equation}
This implies that $\varepsilon$, ${\cal P}$ and $n$ are not independent functions, but they are determined in terms of a single {\it master} function, $P(a,b)$, which is the pressure. In addition, we can identify the local value of the temperature and chemical potential with $a$ and $b$ respectively. 

Let us discuss now the properties of the equilibrium partition function at first order in the derivative expansion. The most general expression compatible with the symmetries mentioned above is~\cite{Banerjee:2012iz,Megias:2014mba}:
\begin{equation}
\hspace{-2cm} \cW_{(1)} = \int d^3x \sqrt{g_3} \left[ \alpha_1(\sigma,A_0) \epsilon^{ijk}A_i A_{jk} + \alpha_2(\sigma,A_0) \epsilon^{ijk} A_i f_{jk} \!+\! \alpha_3(\sigma,A_0) \epsilon^{ijk} a_i f_{jk}  \right] \,, \label{eq:W1}
\end{equation}
where $A_{ij} = \partial_i A_j - \partial_j A_i$ and $f_{ij} = \partial_i a_j - \partial_j a_i$. The coefficients $\alpha_i(\sigma,A_0)$ depend on the particular theory considered, and they can be determined for instance by inserting Eq.~(\ref{eq:W1}) into Eqs.~(\ref{eq:crJ})-(\ref{eq:crT}), and comparing the result with the constitutive relations for that theory. In an ideal gas of Dirac fermions one finds 
\begin{equation}
\hspace{-1.5cm} \alpha_1(\sigma,A_0) =  - \frac{C}{6} \frac{A_0}{T_0} \,, \quad \alpha_2(\sigma,A_0) =  -\frac{1}{2}\left( \frac{C}{6} \frac{A_0^2}{T_0^2} - C_2 \right) \,, \quad \alpha_3(\sigma,A_0) = 0 \,,
\end{equation}
where the coefficients $C = 1/(4\pi^2)$ and $C_2 = 1/24$ are related to the axial anomaly~\cite{Son:2009tf,Erdmenger:2008rm} and mixed gauge-gravitational anomaly~\cite{Landsteiner:2011cp}, respectively. These coefficients induce some contributions to the chiral magnetic and vortical conductivities, which read
\begin{equation}
\sigma_B = C \mu \,, \qquad \sigma_V = \frac{1}{2} C \mu^2 + C_2 T^2 \mu \,,
\end{equation}
where $T = e^{-\sigma} T_0$ and $\mu = e^{-\sigma} A_0$. In the following we will use this formalism that relates the partition function with the constitutive relations of the theory. $\cW$ will be obtained by solving the anomaly equations.

\section{Non-abelian anomalies}
\label{sec:non_abelian}

In this section we will provide a short introduction to chiral anomalies, and study the partition function for non-abelian theories. 

\subsection{The chiral anomaly}
\label{subsec:chiral_anomaly}

Let us consider the theory of a chiral fermion coupled to an external gauge field $\cA_\mu \equiv \cA_\mu^a t_a$ described by the Lagrangian
\begin{equation}
{\mathcal L}_{\YM} = i \overline\psi \gamma^\mu (\partial_\mu - i t_a \cA_\mu^a) \psi \,,
\end{equation}
where $t_a = t_a^\dagger$ are the Hermitian generators of the Lie algebra. To study gauge anomalies it is convenient to work with the effective action functional obtained by integrating out the fermion field
\begin{equation}
e^{i\Gamma[\cA]} \equiv \int {\mathcal D}\overline\psi {\mathcal D}\psi \, e^{i S_{\textrm{\tiny YM}}[\cA,\psi,\overline\psi]} \,.
\end{equation}
Under a general shift $\mcA_\mu^a \to \mcA_\mu^a + \delta \mcA_\mu^a$, the variation of $\Gamma[\cA]$ can be expressed as
\begin{equation}
\delta \Gamma[\mcA] = \int d^4 x \, \delta \mcA_\mu^a(x) \, J_{a\, \cons}^\mu(x) \,, \label{eq:deltaGamma}
\end{equation}
where $J_{a\, \cons}^\mu(x)$ is the consistent current. The axial anomaly is given by the failure of the effective action to be invariant under axial gauge transformations
\begin{equation}
\mcA_\mu  \longrightarrow g^{-1} \mcA_\mu g - i g^{-1} \partial_\mu g  \,, \qquad g(x) = \exp\left( - i \Lambda_a^{\A}(x) t_a \right)  \,.
\end{equation}
Under such a transformation
\begin{equation}
\delta_{\gauge} \Gamma[\mcA]= - \int d^4 x \, \Lambda_a^{\A}(x)  \, G_a[\mcA(x)]  \,,
\end{equation}
where $G_a[\mcA(x)]$ is the consistent anomaly. Particularizing Eq.~(\ref{eq:deltaGamma}) to $\delta \mcA_\mu^a = (D_\mu\Lambda^{\A})^a$, one finds the (non)-conservation law for the consistent current
\begin{equation}
D_\mu J_{a \, \cons}^\mu(x) = G_a[\mcA(x)] \,.
\end{equation}
The consequences of the anomaly to the hydrodynamics of fluids will be analyzed in the rest of the manuscript.

\subsection{The Bardeen form of the anomaly}
\label{subsec:Bardeen_form}

In the following we will consider a non-abelian theory with symmetry group $\U(N_f) \times \U(N_f)$, described by the Lagrangian
\begin{equation}
{\mathcal L}_{\YM} = i \overline\psi_\LL \gamma^\mu (\partial_\mu - i t_a \cA_{\LL\,\mu}^a) \psi_\LL +  i \overline\psi_\RR \gamma^\mu (\partial_\mu - i t_a \cA_{\RR\,\mu}^a) \psi_\RR  \,.
\end{equation}
The Bardeen form of the anomaly in this theory is~\cite{Bardeen:1969md}
\begin{eqnarray}
G_a[\mathcal{V}, \mathcal{A}] &=& \frac{i N_c}{16 \pi^2} \epsilon^{\mu \nu  \alpha \beta} \Tr \Bigl\{ t_a \bigl[ {\mcV_{\mu \nu} \mcV_{\alpha \beta} + \frac{1}{3} \mcA_{\mu \nu} \mcA_{\alpha \beta}}  - {\frac{32}{3} \mcA_\mu \mcA_\nu \mcA_\alpha \mcA_\beta }  \\ 
&& \quad + {\frac{8}{3} i  (\mcA_\mu \mcA_\nu \mcV_{\alpha \beta}  + 
\mcA_\mu \mcV_{\alpha \beta} \mcA_\nu  + \mcV_{\alpha \beta} \mcA_\mu \mcA_\nu) \bigr]} \Bigr\} \,,
\end{eqnarray}
where
\begin{eqnarray}
\mcV_{\mu \nu} &=&  \partial_\mu \mcV_\nu - \partial_\nu \mcV_\mu - i[\mcV_\mu, \mcV_\nu] - i[\mcA_\mu, \mcA_\nu]  \,,  \\ 
\mcA_{\mu \nu} &=&  \partial_\mu \mcA_\nu - \partial_\nu \mcA_\mu - i[\mcV_\mu, \mcA_\nu] - i[\mcA_\mu, \mcV_\nu] \,,
\end{eqnarray}
are the field strengths for the vector and axial gauge fields, and $N_c$ is the number of colors~\footnote{We define the vector and axial gauge fields $(\mcV,\mcA)$ in terms of $(\cA_{\textrm{\tiny L}},\cA_{\textrm{\tiny R}})$ by $\cA_{\textrm{\tiny L}} \equiv \mcV - \mcA$ and $\cA_{\textrm{\tiny R}} \equiv \mcV + \mcA$. Consequently, the corresponding vector and axial components of the currents are related to their left- and right-handed components as $J_{\textrm{\tiny R}}^\mu = \frac{1}{2}\left(J_{\textrm{\tiny V}}^\mu + J_{\textrm{\tiny A}}^\mu\right)$ and $J_{\textrm{\tiny L}}^\mu = \frac{1}{2}\left(J_{\textrm{\tiny V}}^\mu - J_{\textrm{\tiny A}}^\mu\right)$.}. $G_a$ includes triangle, square and pentagon one-loop diagram contributions. The anomaly arises from the breaking of gauge invariance under axial gauge transformations of the effective action~$\Gamma_0[\mcV,\mcA]$, so that the action should satisfy
\begin{equation}
\mathscr{Y}_a(x) \Gamma_0[\mcV, \mcA] = 0 \,, \qquad  \mathscr{X}_a(x) \Gamma_0[\mcV, \mcA] = G_a[\mcV, \mcA] \,,  \label{eq:eq_anomaly}
\end{equation}
where $\mathscr{Y}_a(x)$ and $\mathscr{X}_a(x)$ are the local generators of vector and axial gauge transformations, respectively. The computation of $\Gamma_0[\mcV, \mcA]$ can be performed by solving Eq.~(\ref{eq:eq_anomaly}), leading to
\begin{eqnarray}
\hspace{-1.5cm} \Gamma_0[V, A, G] &=& -\frac{N_c}{32 \pi^2}  \int dt \,  d^3 x  \sqrt{g_3} \, 
 \epsilon^{i j k} \, \Tr \Biggl\{ \frac{32}{3} i \, V_0 A_i A_j A_k  \nonumber \\  
\hspace{-1.5cm}&& \quad + \frac{4}{3} (A_0 A_i + A_i A_0) A_{j k} + 4  (V_0 A_i + A_i V_0) V_{j k} \nonumber \\ 
\hspace{-1.5cm}&& \quad + \frac{8}{3} \bigl(A_0^2 + 3 V_0^2 \bigr) A_i \partial_j a_k \Biggr\} + C_2 T_0^2  \int dt \, d^3 x \sqrt{g_3} \, \epsilon^{i j k}  \Tr A_i \, \partial_j a_k \,, \label{eq:Gamma_0}
\end{eqnarray}
where $V_\mu$ and $A_\mu$ are KK invariant fields. $\Gamma_0[\mcV,\mcA]$ can be determined also from differential geometry methods, cf. Refs.~\cite{Jensen:2013rga,Manes:2018llx,Manes:2019fyw,Manes:2020zdd} for details. As it is mentioned in Sec.~\ref{subsec:derivative_expansion}, the coefficient $C_2$ is related to the mixed gauge-gravitational anomaly. While in principle it would be possible to compute this contribution by taking into account the Riemann tensor effects, in the following we will neglect it as this analysis goes beyond the scope of the present work.

\section{Covariant currents and constitutive relations without Nambu-Goldstone bosons}
\label{sec:cov_currents}

In this section we will study the constitutive relations with unbroken chiral symmetry. We will particularize the result for a background with two light quark flavors.

\subsection{Covariant currents}
\label{subsec:cov_currents}

The charged currents obtained from the functional derivatives of the effective action are consistent currents, cf. Eq.~(\ref{eq:Jcr}).  However, only covariant currents can enter in the constitutive relations, and these cannot be obtained directly from the functional derivative of an effective action, but they are defined by adding to the consistent currents the Bardeen-Zumino (BZ) polynomials~\cite{Bardeen:1984pm}, i.e.
\begin{equation}
J^\mu_\cov = J^\mu_\cons + J^\mu_{\BZ} \,,  \label{eq:Jcov_Jcons_JBZ}
\end{equation}
with~\cite{Manes:2018llx,Manes:2019fyw}
\begin{eqnarray}
J_{\V \, \BZ}^\mu &=& - \frac{N_c}{8 \pi^2} \epsilon^{\mu \nu \alpha \beta} 
\Tr \Bigr\{ t_a \bigl(\mathcal{A}_\nu \mathcal{V}_{\alpha \beta} + \mathcal{V}_{\nu \alpha} \mathcal{A}_\beta + \frac{8}{3} i \, \mathcal{A}_\nu  \mathcal{A}_\alpha \mathcal{A}_\beta  \bigr)\Bigr\} \,, \label{eq:JBZ_V}  \\
J_{\A \, \BZ}^\mu &=& - \frac{N_c}{24 \pi^2} \epsilon^{\mu \nu \alpha \beta} 
\Tr  \Bigr\{ t_a \left(\mathcal{A}_\nu \mathcal{A}_{\alpha \beta} +
        \mathcal{A}_{\nu \alpha} \mathcal{A}_\beta  \right) \Bigr\} \,.  \label{eq:JBZ_A}
\end{eqnarray}
Then the covariant currents and energy-momentum tensor in equilibrium can be obtained from ${\mathcal W}_0 = i\Gamma_0$ by using Eqs.~(\ref{eq:Jcr}), (\ref{eq:Tcr}) and (\ref{eq:Jcov_Jcons_JBZ}). The result is~\cite{Manes:2019fyw}
\begin{eqnarray} 
\langle J^i_{a\, \V} \rangle_{\cov}  &=&  
 \frac{N_c}{8 \pi^2} e^{-\sigma} \epsilon^{i j k} \Tr \Bigl\{ t_a \bigl[ 
   (A_0 V_{j k} +  V_{j k} A_0)  + 
   (V_0 A_{j k} +  A_{j k} V_0) \nonumber \\ 
&&\quad + 2 (A_0 V_0 +V_0 A_0)  \partial_j a_k  \bigr]   \Bigr\}, \label{eq:Ji_cov}\\ 
\langle J^i_{a \, \A} \rangle_{\cov}  &=& \frac{N_c}{8 \pi^2} e^{-\sigma} \epsilon^{i j k} \Tr \Bigl\{ t_a \bigl[ (A_0 A_{j k} +  A_{j k} A_0) + (V_0 V_{j k} +  V_{j k} V_0) \nonumber \\ 
&&\quad + 2 (A_0^2 +V_0^2)  \partial_j a_k  \bigr]   \Bigr\} \,, \label{eq:Ji5_cov} \\
\langle T_0\,{}^i \rangle &=& -\frac{N_c}{8\pi^2} e^{-\sigma}  \epsilon^{i j k} \, \Tr \Bigl\{ \bigl(A_0^2 + V_0^2 \bigr) A_{j k} + (V_0 A_0 + V_0 A_0) V_{j k}   \nonumber \\ 
&&\quad +  \biggl( \frac{2}{3}A_0 ^3 + 2 A_0 V_0^2  \biggr)  \partial_j a_k  \Bigr\}  \,,  \label{eq:T0i_cov}
\end{eqnarray}
and vanishing values for the time components of the currents $\langle J_{0 \, a \, \V} \rangle_{\cov} = \langle J_{0 \, a \, \A} \rangle_{\cov} = 0$, and for the other components of the energy-momentum tensor $\langle T_{00} \rangle = \langle T^{ij} \rangle = 0$.

While these expressions are valid for a non-abelian theory with symmetry group $\U(N_f) \times \U(N_f)$, we will particularize them for a specific background. In presence of non-abelian charges, the maximal number of chemical potentials to be consistently introduced corresponds to the dimension of the Cartan subalgebra. Let us consider the following background for $N_f = 2$
\begin{equation}
V_\mu(\vec{x}) = V_{\mu \, 0}(\vec{x})  t_0 + \, V_{\mu \, 3}(\vec{x}) t_3 \,, \qquad A_0 = A_{0 \, 0} \, t_0 \,, \qquad A_i = 0 \,, \label{eq:background}
\end{equation}
where  $t_0 = \frac{1}{2}  1_{2\times 2}$ and $t_3 = \frac{1}{2} \sigma_3$, while $\sigma_i$ are the Pauli matrices~\footnote{Notice that $A_i=0$ does not imply a vanishing spatial component of the gauge field, as this turns out to be $\mathcal{A}_i = a_i A_{0 \, 0} \, t_0$ by Eq.~(\ref{eq:A_KK}).}. In the following we will consider that $A_{0 \, 0}$ is constant. In addition, we can define the equilibrium velocity field, as well as equilibrium baryonic, isospin and axial chemical potentials, as
\begin{equation}
u_\mu = -e^\sigma (1, a_i) \,, \quad \mu_0 = e^{-\sigma} \, V_{0 \, 0}   \,, \quad \mu_3 =  e^{-\sigma} \, V_{0 \, 3}  \,, \quad  \mu_5 =  e^{-\sigma} \, A_{0 \, 0}  \,,
\end{equation}
respectively, where $\mu_5$ controls the chiral imbalance of the system~\cite{Gatto:2011wc,Andrianov:2013qta}. The coupling to the external gauge fields comes through the magnetic components of the non-abelian vector field strength, while the explicit dependence on $u_\mu$ is codified in terms of the vorticity vector. These quantities are defined by
\begin{equation}
{\mathcal B}_a^\mu = \frac{1}{2} \epsilon^{\mu\nu\alpha\beta} u_\nu \mathcal{V}_{\alpha\beta \, a} \,, \qquad \omega^\mu = \frac{1}{2} \epsilon^{\mu\nu\alpha\beta} u_\nu \partial_\alpha u_\beta \,.
\end{equation}
Then, the constitutive relations can be computed by using Eqs.~(\ref{eq:Ji_cov})-(\ref{eq:T0i_cov}), and the result expressed in Lorentz covariant form is
\begin{eqnarray}
\langle J_{0 \, \V}^\mu \rangle_{\cov} &=& \frac{N_c }{8 \pi^2} \mu_5 {\mathcal B}_0^\mu  , \qquad \langle J_{3\, \V}^\mu \rangle_{\cov} =  \frac{ N_c }{8 \pi^2} \mu_5 {\mathcal B}_3^\mu  \,, \\
\langle J_{0\, \A}^\mu \rangle_{\cov} &=& \frac{N_c}{8 \pi^2}  \Big(  \mu_0 {\mathcal B}_0^\mu  + \mu_3  {\mathcal B}_3^\mu  + (\mu_0^2 + \mu_3^2 - \mu_5^2) \omega^\mu \Big) \, ,  \\ 
\langle J_{3\, \A}^\mu \rangle_{\cov} &=& \frac{N_c}{8 \pi^2} \Big( \mu_3 {\mathcal B}_0^\mu  + \mu_0 {\mathcal B}_3^\mu  + 2 \mu_0 \mu_3  \omega^\mu \Big) \,,
\end{eqnarray}
for the currents, and
\begin{equation}
\hspace{-2.2cm} \langle T^{\mu\nu} \rangle  =  u^\mu q^\nu + u^\nu q^\mu \quad \textrm{with} \quad q^\mu = \frac{N_c}{8 \pi^2} \mu_5 \bigg[  \mu_0 {\mathcal B}_0^\mu  +  \mu_3 {\mathcal B}_3^\mu +  \left(\mu_0^2 + \mu_3^2 - \frac{1}{3}\mu_5^2\right) \omega^\mu \bigg] \,,
\end{equation}
for the energy-momentum tensor.

\subsection{Electromagnetic, baryon and isospin currents}
\label{subsec:em_bar_iso_currents}

Let $\psi$ be a flavor doublet of Dirac spinors made out of up and down quarks, i.e. $\psi = \left( \begin{array}{c}
u    \\
d
\end{array} \right)$. Electromagnetism is identified with the $\U(1)_\V$ subgroup defined by the charge matrix for two light flavors with electric charges $+\frac{2}{3}e$ and $-\frac{1}{3}e$, i.e.
\begin{equation}
Q = \left(
\begin{array}{cc}
\frac{2}{3} &          0         \\
       0    &  -\frac{1}{3}  
\end{array} \right)  = \frac{1}{3}t_0 + t_3 \,. \label{eq:Q}
\end{equation}
Then, using that $J_{a\, V \, \cons}^\mu = \overline\psi \gamma^\mu t_a \psi$, we can distinguish between the electromagnetic, baryonic and isospin currents, defined as
\begin{eqnarray}
J^\mu_{\eem \, \cons} &=& e \overline\psi \gamma^\mu Q \psi = \frac{e}{3} J_{0\, \V \, \cons}^\mu  + e J_{3\, \V \, \cons}^\mu \,, \nonumber \\
J^\mu_{\bbar \, \cons} &=& \frac{2}{3} J_{0\, \V \, \cons}^\mu \,, \nonumber \\
J^\mu_{\iiso \, \cons} &=& J_{3 \, \V \, \cons}^\mu \,,
\end{eqnarray}
respectively. These currents are not independent, but they fulfill the Gell-Mann-Nishijima (GMN) relation $J^\mu_{\eem \, \cons} = \frac{e}{2} J^\mu_{\bbar \, \cons} + e  J^\mu_{\iiso \, \cons}$. The same pattern is followed by the BZ terms of each current, and then the same relations are satisfied by the corresponding covariant currents. If we denote the physical magnetic field by  ${\mathcal B}^\mu = \frac{1}{2} \epsilon^{\mu\nu\alpha\beta} u_\nu \mbhV_{\alpha\beta}$ where the physical potential is $\mbhV_\mu$ and its KK invariant form is $\mbbV_\mu$, i.e. $\mbbV_0 = \mbhV_0$ and $\mbbV_i = \mbhV_i - a_i \mbhV_0$, then after making the replacements $V_{\mu\,0} = \frac{e}{3} \mbbV_\mu$ and $V_{\mu\,3} = e \mbbV_\mu$, one finds 
\begin{equation}
\langle  J^\mu_{\eem} \rangle_{\cov} = \frac{5 e^2 N_c}{36 \pi^2} \mu_5 {\mathcal B}^\mu \,,
\end{equation}
where we have used that ${\mathcal B}_3^\mu = 3 {\mathcal B}_0^\mu = e {\mathcal B}^\mu$. This expression gives the transport coefficient associated with the CME. The absence of a CVE in the vector currents is a direct consequence of having considered the flavor group $\U(2)_\V \times \U(2)_\A$. This state of affairs contrasts with the $\U(1)_\V \times \U(1)_\A$ case, studied in~\cite{Landsteiner:2012kd,Jensen:2013vta}, where the cancellation leading to a vanishing value for the chiral vortical conductivity does not take place.

\section{Effective action in presence of spontaneous symmetry breaking: the Wess-Zumino-Witten partition function}
\label{sec:SBB}

In this section we will consider the physical situation in which the symmetry is spontaneously broken, either total or partially. A consequence is the appearance of NG bosons that can couple to external gauge fields and contribute to the anomaly. The WZW partition function describes the effects of the anomaly when the symmetry is spontaneously broken, and accounts for the anomaly-induced interactions between the external gauge fields $\mcA$ and the NG bosons~$\xi^a$~\cite{Kaiser:2000ck,Son:2007ny,Fukushima:2012fg,Brauner:2017mui}. The WZW action admits a simple expression in terms of the anomalous functional in absence of symmetry breaking $\Gamma_0$ studied in Sec.~\ref{sec:non_abelian}~\cite{Wess:1971yu,Witten:1983tw,Manes:1984gk,Chu:1996fr}
\begin{equation}
\Gamma^{\WZW}[\mathcal A, \xi] =  \Gamma_0[\mathcal A] - \Gamma_0[\mathcal A_{g}] \,, \label{eq:WZW}
\end{equation}
where $\mathcal A_{g} = g^{-1} \mathcal A g + g^{-1} dg$ is the gauge transformed field with group element $g \equiv \exp(-i \xi^a t_a)$. For applications to hadronic fluids, we are interested in the case $U(2)_\LL \times U(2)_\RR \to U(2)_\V$, where the symmetry is broken down to the diagonal subgroup of vector gauge transformations. Then, one can make the replacements $\mathcal A \to (\mathcal A_\LL, \mathcal A_\RR)$ and $g \to (U,{\mathbb I})$, where ${\mathbb I}$ is the identity element and
\begin{equation}
\hspace{-1.5cm} U(\xi) = \exp \Bigl( i X(\xi) \Bigr) \quad \textrm{with} \quad X(\xi) := 2\sum_{a=1}^3 \xi_a t_a = \frac{\sqrt{2}}{f_\pi} \left(
\begin{array}{cc}
\frac{1}{\sqrt{2}} \pi^0 & \pi^+ \\
\pi^- & -\frac{1}{\sqrt{2}} \pi^0
\end{array}  \right)\,, 
\end{equation}
includes three NG bosons $\{ \pi^0, \pi^\pm \}$ from the broken $\SU(2)_\A$ symmetry, while $f_\pi \approx 92 \, \MeV$ is the pion decay constant~\footnote{The fourth NG boson $\xi_0$ is absent, as the $\U(1)_{\textrm{\tiny A}}$ symmetry is violated by non-perturbative effects.}. Then the action at the lowest order in derivatives can be written as 
\begin{equation}
\mathcal W_{(0)} = \frac{1}{T_0} \int d^3x \sqrt{g_3} \, e^\sigma \left[ P(T,\mu_0,\mu_3)  + \mathcal L\right] \,, \label{eq:W0}
\end{equation}
where $P$ is the pressure in absence of NG bosons already introduced in Sec.~\ref{subsec:derivative_expansion}, and the Lagrangian contains the dependence on the pions
\begin{equation}
\mathcal{L} = \frac{f_\pi^2}{4} G^{\mu\nu} \Tr \left\{ D_\mu U (D_\nu U)^\dagger \right\}  \,.
\end{equation}
At first order in derivatives, the correction to the partition function is given by the WZW action evaluated in the background of Eq.~(\ref{eq:background}). The WZW action can be computed following the prescription of Eq.~(\ref{eq:WZW}), and the result is
\begin{eqnarray}
\hspace{-2.4cm} \mathcal W_{(1)}^{\WZW} &=& \frac{N_c}{8 \pi^2 T_0}  \int d^3 x \sqrt{g_3}  \, \epsilon^{i j k} \Bigg[
- \frac{1}{2}  V_{0 \, 0}  V_{i \, 3} \partial_j \Tr \bigl\{ (R_k + L_k) Q \bigr\} + \frac{i}{6} \,  V_{0 \, 0} \Tr \bigl\{ L_i L_j L_k  \bigr\}  \nonumber \\ 
\hspace{-2.4cm}&+&  \frac{1}{2} \left( V_{0 \, 0} \,  \partial_i V_{j \, 3} +  V_{0 \, 3} \, \partial_i V_{j \, 0}  +  \frac{1}{2} V_{0 \, 0}  V_{0 \, 3} \, f_{ij} \right) \Tr \bigl\{ (R_k +  L_k) Q \bigr\}  \label{eq:W_WZW} \\ 
\hspace{-2.4cm}&+& \frac{1}{6}  A_{0 \, 0} \left( \partial_i V_{j \, 3} +  \frac{1}{2} V_{0 \, 3} f_{ij} \right) \left( \Tr \bigl\{ (R_k -L_k) Q \bigr\}  - 2 V_{k \, 3}  \Tr  \bigl\{Q (Q - U^{-1} Q U) \bigl\}  \right)  \Bigg] \,, \nonumber
\end{eqnarray}
where we have introduced the notation
\begin{equation}
L_j = i \partial_j U \, U^{-1} \quad \textrm{and} \quad R_j = i U^{-1} \partial_j U \,.
\end{equation}
Since $U$ takes values on $\SU(2)$, the generator $t_3$ can be interchanged with the charge matrix $Q$ inside the traces.

\section{Constitutive relations of the two-flavor hadronic fluid}
\label{subsec:constitutive_relations}

Physical quantities in hydrodynamics admit the decomposition in terms of their PF contributions and corrections containing higher terms in derivatives, i.e.
\begin{equation}
J^\mu = J^\mu_{\PPFF} + \delta J^\mu  \,, \qquad T^{\mu\nu} = T^{\mu\nu}_{\PPFF} +  \pi^{\mu\nu} \,.
\end{equation}
These corrections can be either dissipative or anomalous. Different definitions of the same physical variable may vary by gradient-dependent terms $(\delta T, \delta \mu_a, \delta u^\mu, \dots)$. This leads to high order terms ambiguities that should be compensated by the PF constitutive relations
\begin{equation}
J^\mu_{\PPFF}(T_0 + \delta T, \mu_{a\, 0} + \delta \mu_a, \dots)  \,, \qquad T^{\mu\nu}_{\PPFF}(T_0 + \delta T, \mu_{a\, 0} + \delta \mu_a, \dots) \,,
\end{equation} 
as the form of the currents and energy-momentum tensor cannot be changed by the ambiguities. The particular frame to fix these ambiguities is chosen in the following by requiring that one-derivative corrections to PF quantities vanish, so that contributions at this order come only from the terms $\delta J^\mu$ and $\pi^{\mu\nu}$. Using this frame, it was found that for systems with spontaneous symmetry breaking the energy-momentum tensor receives no corrections, while the corrections to the charged currents admit a decomposition in terms of their longitudinal and transverse components~\cite{Manes:2018llx,Manes:2019fyw}, i.e.
\begin{equation}
\pi^{\mu\nu} = 0 \,, \qquad \delta J^\mu = -  (u_\nu \delta J^\nu) u^\mu + P^\mu{}_\nu \delta J^\nu \,, 
\end{equation} 
where $P^{\mu\nu} = G^{\mu\nu} + u^\mu u^\nu$ is the transverse projector to the local fluid velocity.

\subsection{Constitutive relations at the lowest order}
\label{subsec:CR_0}

The constitutive relations at the lowest order can be obtained by taking the corresponding functional derivatives on the effective action of Eq.~(\ref{eq:W0}), cf. Eqs.~(\ref{eq:Jcr}) and (\ref{eq:Tcr}). This leads to the result
\begin{eqnarray}
\hspace{-2.3cm} \langle J_{\mu\, 0} \rangle_{\PPFF} &=& n_0 \, u_\mu  \,, \\
\hspace{-2.3cm} \langle J_{\mu\, 3} \rangle_{\PPFF} &=& n_3 \, u_\mu  + i \frac{f_\pi^2}{4} \Tr  \left\{ [Q,U] \partial_\mu U^\dagger + [Q,U^\dagger] \partial_\mu U \right\} + \frac{f_\pi^2}{2} V_{\mu \, 3}   \Tr  \left\{ [Q,U] [Q,U^\dagger] \right\} \,,  \\
\hspace{-2.3cm} \langle T^{\mu\nu} \rangle_{\PPFF} &=& (\varepsilon + P) u^\mu u^\nu + P G^{\mu\nu} +  { \frac{f_\pi^2}{4} G^{\mu\alpha} G^{\nu\beta} \Tr\left\{ D_{\alpha} U ( D_{\beta} U)^\dagger +  D_{\beta} U ( D_{\alpha} U)^\dagger \right\} } \,,
\end{eqnarray}
where the number densities are defined by $n_a = \partial P/\partial \mu_a$ $(a = 0,3)$. These contributions to the constitutive relations have been expressed in a covariant form by writing them in terms of the metric $G_{\mu\nu}$ and the four velocity $u^\mu$, cf. Eqs.~(\ref{eq:metric}) and (\ref{eq:umu_P}). Notice that since the BZ terms contain one derivative of the gauge fields, there is no distinction between consistent and covariant currents at leading order in the derivative expansion.

\subsection{Corrections to the leading order constitutive relations}
\label{subsec:CR_1}

All dependence on the NG bosons matrix $U$ in the constitutive relations comes in terms of the following covariant expressions
\begin{eqnarray}
\mathcal{H} &=& \Tr \bigl\{ \left( U^{-1} Q U  - Q\right) Q \bigr\} \,,  \qquad \mcI_\mu  \equiv \Tr \bigl\{ (R_\mu + L_\mu) Q \bigr\} \,, \label{eq:HI} \\
\mcT_\mu  &\equiv&  \Tr \bigl\{ (R_\mu - L_\mu) Q \bigr\} + 2 \mathcal{V}_{\mu \, 3}  \Tr \bigl\{ \left( U^{-1} Q U  - Q \right)  Q \bigr\} \,. \label{eq:Tmu}
\end{eqnarray}
Then, the currents can be decomposed into their longitudinal and transverse components, and the result can be written as linear combinations of the following five pseudo-scalar quantities
\begin{eqnarray}
\hspace{-2cm} \mathbb S_{1(a)} &\equiv& \mcI_\mu \, {\mathcal B}_a^\mu  \,, \qquad \mathbb S_2 \equiv \mcI_\mu \, \omega^\mu  \,, \qquad \mathbb S_3 \equiv   \epsilon^{\mu\nu\alpha\beta}  u_\mu \left[ \mathcal{V}_{\nu \, 3} \, \partial_\alpha \mcI_\beta - \frac{i}{3} \Tr\{ L_\nu L_\alpha L_\beta \} \right] \,, \\
\hspace{-2cm} \mathbb S_{4 (a)} &\equiv& \mcT_\mu \, {\mathcal B}_a^\mu   \,, \qquad \mathbb S_5 \equiv  \mcT_\mu \, \omega^\mu  \,,
\end{eqnarray}
and the four transverse pseudo-vectors
\begin{eqnarray}
P_{1(a)}^\mu &=& \frac{1}{T} \epsilon^{\mu\nu\alpha\beta} u_\nu \mcI_\alpha \, \mathcal E_{\beta(a)} \,,  \qquad P_2^\mu =  \epsilon^{\mu\nu\alpha\beta} u_\nu \partial_\alpha \mcI_\beta  \,,  \\
P_{3(a)}^\mu &=& \frac{1}{T}\epsilon^{\mu\nu\alpha\beta} u_\nu \mcT_\alpha \, \mathcal E_{\beta(a)} \,, \qquad P_{4}^\mu = \epsilon^{\mu\nu\alpha\beta} u_\nu \partial_\alpha \mcT_\beta \,,
\end{eqnarray}
where $\mathcal E_{\mu(a)} = \mathcal V_{\mu\nu\, a} \, u^\nu = T \partial_\mu\left( \mu_a/T \right)$ is the electric field for $a= 0, 3$. Finally, the constitutive relations at first order in derivatives read
\begin{eqnarray}
\hspace{-1cm} u_\mu \langle \delta J_{0\, \V}^\mu \rangle_{\cov} &=& -\frac{N_c}{16\pi^2} \left[ \mathbb S_{1(3)} + \mathbb S_{3} \right]   \,,  \\
\hspace{-1cm} P^\mu{}_\nu \langle \delta J^\nu_{0\,\V} \rangle_{\cov} &=& - \frac{N_c}{16\pi^2} \left[ T \mathbb P_{1(3)}^\mu - \mu_3 \mathbb P_2^\mu - 2 \mu_5 \mathcal B_0^\mu \right] \,,\\
\hspace{-1cm} u_\mu \langle\delta J_{3\, \V}^\mu \rangle_{\cov} &=&  - \frac{N_c}{48\pi^2} \left[ 3 \mathbb S_{1(0)} + 2 \mu_5 \mathbb S_{5} \right] \,,  \\
\hspace{-1cm} P^\mu{}_\nu \langle \delta J^\nu_{3\, \V} \rangle_{\cov} &=&  - \frac{N_c}{48\pi^2} \left[ 3 T \mathbb P_{1(0)}^\mu + \mu_5  \mathbb P_{4}^\mu + 4 \mu_3 \mu_5 \mathcal H \omega^\mu - 2 \mu_5 (\mathcal H + 3) \mathcal B_3^\mu \right] \,, 
\end{eqnarray}
for vector currents, and
\begin{eqnarray}
u_\mu \langle \delta J_{0\, \A}^\mu \rangle_{\cov} &=&  - \frac{N_c}{48\pi^2} \mathbb S_{4(3)} \,, \\
P^\mu{}_\nu \langle \delta J^\nu_{0 \, \A} \rangle_{\cov} &=& -\frac{N_c}{48\pi^2} \left[ T \mathbb P_{3(3)}^\mu + 2 \mu_3 {\mathcal H} \mathcal B_3^\mu + 4\mu_5^2 \omega^\mu \right]  \,, \\
u_\mu \langle \delta J_{3\, \A}^\mu \rangle_{\cov} &=&  -\frac{N_c}{48\pi^2} \left[ 3 \mathbb S_{4(0)} - 2 \mu_5 \mathbb S_2 \right] \,, \\
P^\mu{}_\nu \langle \delta J^\nu_{3 \, \A} \rangle_{\cov} &=& -\frac{N_c}{48\pi^2} \left[ 3 T P_{3(0)}^\mu + 6 \mu_3 {\mathcal H} {\mathcal B}_0^\mu - \mu_5 \mathbb P_2^\mu  \right] \,,
\end{eqnarray}
for axial-vector currents. The BZ contributions, which correspond to the CME and CVE, are those terms proportional to the magnetic field ${\mathcal B}_a^\mu$ and vorticity vector $\omega^\mu$ without gothic fonts prefactors, and they only appear in the transverse components of the currents. Notice as well that the covariant currents are given in terms of the KK-variant gauge fields $(\mcV_{\mu \, a }, \mcA_{\mu\, a})$ without any explicit reference to the KK gauge field~$a_i$. Finally, the terms proportional to $\mathbb P_{1(a)}^\mu$ and $\mathbb P_{3(a)}^\mu$ in the constitutive relations are the ones associated to the CEE, i.e. charge transport normal to the direction of the electric field~\cite{Neiman:2011mj}. To compare our results with other analyses in the literature we expand the covariant expressions of Eqs.~(\ref{eq:HI}) and (\ref{eq:Tmu}) in powers of the pion fields. Then, using the definitions of the currents and electromagnetic field given in Sec.~\ref{subsec:em_bar_iso_currents}, we can express the electromagnetic current in terms of the pion fields, electric and magnetic fields as
\begin{equation}
\hspace{-1.7cm} \langle \delta J^i_{\eem} \rangle_{\cov} =  \frac{e^2 N_c}{12\pi^2}  \left[ \frac{1}{f_\pi} \epsilon^{ijk} \partial_j \pi^0 \mathcal E_k + \frac{2}{f_\pi^2} \mu \mu_5 \pi^+ \pi^- \omega^i + \frac{5}{3} \mu_5 {\mathcal B}^i \right] + \cdots  \,,
\end{equation}
where the physical electric field $\mathcal E_i$ and the electric charge chemical potential $\mu$ are defined as
\begin{equation}
\mathcal E_i = e^{-\sigma} \partial_i \mbhV_0 = T \partial_i\left( \frac{\mu}{T} \right) \,, \qquad  \mu \equiv e^{-\sigma} \mbhV_0  \,,
\end{equation}
while the relation with the baryonic and isospin chemical potentials is~$\mu_0 = \frac{1}{3} \mu_3 = \frac{e}{3} \mu$. Similar expressions can be written for the baryon and isospin currents, leading to
\begin{eqnarray}
\hspace{-1.7cm} \langle \delta J^i_{\bbar} \rangle_{\cov} &=&  \frac{e N_c}{12\pi^2} \left[ \frac{1}{f_\pi} \epsilon^{ijk} \partial_j \pi^0 \mathcal E_k + \frac{1}{3} \mu_5 \mathcal B^i  \right] + \cdots  \,, \label{eq:Jibar} \\
\hspace{-1.7cm} \langle \delta J^i_{\iiso} \rangle_{\cov} &=&  \frac{e N_c}{24\pi^2} \left[ \frac{1}{f_\pi} \epsilon^{ijk} \partial_j \pi^0 \mathcal E_k + \frac{4}{f_\pi^2} \mu \mu_5 \pi^+ \pi^- \omega^i + 3 \mu_5 {\mathcal B}^i  \right] + \cdots  \,.  \label{eq:Jiiso}
\end{eqnarray}
All three currents are invariant under the gauge transformations of electromagnetism. Notice also that the terms proportional to the vorticity vector come always multiplied by $\mu_5$, which means that in the absence of chiral imbalance $(\mu_5 = 0)$ there are no contributions depending of the vorticity. Finally, by considering similar steps we can obtain some of the explicit contributions to the axial-vector covariant currents. The result is
\begin{eqnarray}
\hspace{-1.9cm} \langle \delta J^i_{0 \, \A} \rangle_{\cov} &=&  \frac{N_c}{24 \pi^2 f_\pi^2} \left[ i e \epsilon^{ijk} (\pi^- \partial_j \pi^+ - \pi^+ \partial_j \pi^- ) \mathcal E_k + 2 e \mu \pi^+ \pi^- \mathcal B^i  - 2\mu_5^2 \omega^i  \right] + \cdots  \,, \label{eq:Ji0A} \\
\hspace{-1.9cm} \langle \delta J^i_{3 \, \A} \rangle_{\cov} &=&  \frac{N_c}{24 \pi^2 f_\pi^2} \left[ i e \epsilon^{ijk} (\pi^- \partial_j \pi^+ - \pi^+ \partial_j \pi^- ) \mathcal E_k - 2 e \mu \pi^+ \pi^- \mathcal B^i  \right] + \cdots  \,. \label{eq:Ji3A}
\end{eqnarray}
These currents contain chiral separation effect terms of electric, magnetic, and vortical type. Let us mention that written in terms of the KK-invariant magnetic field, $\mathbb B^\mu = \frac{1}{2} \epsilon^{\mu\nu\alpha\beta} u_\nu  \mbbV_{\alpha\beta} = \mathcal B^\mu + \frac{6}{e} \mu_0 \omega^\mu$, we find vorticity dependent terms mediated by the charged pion fields, that survive in the case $\mu_5 = 0$. The emergence  of the CEE, i.e. terms proportional to $\epsilon^{ijk} \mathcal E_k$, is manifest in all the currents of Eqs.~(\ref{eq:Jibar})-(\ref{eq:Ji3A}). Despite the ongoing discussion in the literature concerning the non-dissipative character of the CEE (see e.g. Ref.~\cite{Chapman:2013qpa}), our derivation shows that the CEE is intrinsically non-dissipative, so that it does not lead to entropy production.

As a remark, let us mention that in this work we have computed the covariant currents by performing functional derivatives of the effective action and adding the corresponding BZ currents. However, there is a direct procedure to obtain the covariant currents in presence of spontaneous symmetry breaking by using a direct relation with the BZ currents, thus bypassing the need to use the WZW action. Basically, the covariant currents are given by $J^\mu_{\cov}(\mathcal A,g) = g J^\mu_{\BZ}(\mathcal A_g) g^{-1}$~\cite{Jensen:2013kka,Manes:2018llx}, and one just have to make the replacements  $\mathcal A \to (\mathcal A_\LL, \mathcal A_\RR)$ and $g \to (U,{\mathbb I})$ in this relation to obtain
\begin{equation}
\hspace{-2cm} J^\mu_{\LL\, \cov}(\mathcal A_\LL,\mathcal A_\RR,U) =  U J^\mu_{\LL \, \BZ}(\mathcal A_\LL^U, \mathcal A_\RR) U^{-1}  \,, \quad  J^\mu_{\RR\, \cov}(\mathcal A_\LL,\mathcal A_\RR,U) =  J^\mu_{\RR \, \BZ}(\mathcal A_\LL^U, \mathcal A_R) \,, \label{eq:JcovJBZ}
\end{equation}
where the BZ currents are given by Eqs.~(\ref{eq:JBZ_V}) and (\ref{eq:JBZ_A}). Notice that there is no BZ contribution to the energy-momentum tensor, $T^{\mu\nu}_\BZ = 0$, and this is consistent with the fact that the anomalous energy-momentum tensor must vanish in a system with spontaneously broken symmetry, $\langle \pi^{\mu\nu}\rangle = 0$. This procedure has been studied in detail in Ref.~\cite{Manes:2019fyw}.

\section{Conclusions}
\label{sec:conclusions}

In this work we have studied non-dissipative transport effects of relativistic fluids up to first order in the hydrodynamic expansion in presence of non-abelian anomalies. The computation has been performed by using the equilibrium partition function formalism, which is suitable for the study of non-dissipative effects like the CME and CVE. We have extended the analysis to the case of spontaneous symmetry breaking, leading to relevant information about the hydrodynamics of NG bosons interacting with external electromagnetic fields and in presence of vortices in the fluid. After particularization of the results for two flavors, we have been able to provide explicit expressions for the constitutive relations of the covariant currents in a chiral hadronic fluid, and found that the presence of NG bosons induces the CME and CVE both in the vector and axial currents. The vorticity terms are controlled by the chemical potential governing chiral imbalance, $\mu_5$, and disappear in the limit $\mu_5 \to 0$. Our calculation also predicts the emergence of a CEE whose corresponding transport coefficient is explicitly evaluated. Our findings are in agreement with previous results in the literature in the presence of chiral imbalance. 

Gravitational and/or mixed gauge-gravitational anomalies~\cite{Landsteiner:2011cp,Jensen:2012kj} can also be incorporated into the description by considering appropriate curvature terms within the differential geometry methods, a study that is indeed of interest given their recently discovered experimental signatures~\cite{Gooth:2017mbd}. Finally, let us stress that the techniques presented in this work can be extended to a wide variety of physical situations, ranging from superfluids~\cite{Lin:2011aa,Hoyos:2014nua} to condensed matter systems affected by triangle anomalies~\cite{Basar:2013iaa,Landsteiner:2013sja}.

\ack 

The work of J.L.M. and M.V. has been supported by Spanish Science
Ministry grant PGC2018-094626-B-C21 (MCIU/AEI/FEDER, EU) and Basque
Government grant IT979-16. The research of E.M. is supported by
Spanish Science Ministry grant FIS2017-85053-C2-1-P, by FEDER/Junta de
Andaluc\'{\i}a-Consejer\'{\i}a de Econom\'{\i}a y Conocimiento
2014-2020 Operational Programme grant A-FQM-178-UGR18, by Junta de
Andaluc\'{\i}a grant FQM-225, and by Consejer\'{\i}a de Conocimiento,
Investigaci\'on y Universidad of Junta de Andaluc\'{\i}a and European
Regional Development Fund (ERDF) grant SOMM17/6105/UGR. The research
of E.M. is also supported by the Ram\'on y Cajal Program of the
Spanish Science Ministry grant RYC-2016-20678. M.A.V.-M. acknowledges
the financial support from the Spanish Science Ministry through
research grant PGC2018-094626-B-C22 (MCIU/AEI/FEDER, EU), as well as
from Basque Government grant IT979-16.

\section*{References}

\end{document}